\documentclass[conference]{IEEEtran}
\usepackage[T1]{fontenc}
\usepackage{graphicx}
\usepackage{color}
\usepackage{mathtools}
\usepackage{amsthm}
\usepackage{algorithm}
\usepackage[noend]{algpseudocode}
\usepackage{float}
\usepackage{tabularx}
\usepackage{hyperref}
\usepackage{multirow}
\usepackage{arydshln}

\usepackage{xcolor}
\usepackage{enumitem}
\usepackage{caption}
\definecolor{deepblue}{RGB}{0,0,148}
\newcommand{\EXCTGMORE}{8}
\newcommand{\DYNMORE}{25}
\newcommand{\DYNMORETOEXCTG}{17}
\begin{document}

\title{Extended CTG Generalization and Dynamic Adjustment of Generalization Strategies in IC3}

\author{
    \IEEEauthorblockN{Yuheng Su}
    \IEEEauthorblockA{
        University of Chinese Academy of Sciences\\
        Institute of Software, Chinese Academy of Sciences\\
        gipsyh.icu@gmail.com
    }
    \and
    \IEEEauthorblockN{Qiusong Yang\IEEEauthorrefmark{1}}
    \IEEEauthorblockA{
        Institute of Software, Chinese Academy of Sciences\\
        qiusong@iscas.ac.cn
    }
    \and
    \IEEEauthorblockN{Yiwei Ci}
    \IEEEauthorblockA{
        Institute of Software, Chinese Academy of Sciences\\
        yiwei@iscas.ac.cn
    }
    \and
    \IEEEauthorblockN{Ziyu Huang}
    \IEEEauthorblockA{
        Beijing Forestry University\\
        fyy0007@bjfu.edu.cn
    }
}

\maketitle
\begin{abstract}
The IC3 algorithm is widely used in hardware formal verification, with generalization as a crucial step. Standard generalization expands a cube by dropping literals to include more unreachable states. The CTG approach enhances this by blocking counterexamples to generalization (CTG) when dropping literals fails. In this paper, we extend the CTG method (EXCTG) to put more effort into generalization. If blocking the CTG fails, EXCTG attempts to block its predecessors, aiming for better generalization. While CTG and EXCTG offer better generalization results, they also come with increased computational overhead. Finding an appropriate balance between generalization quality and computational overhead is challenging with a static strategy. We propose DynAMic, a method that dynamically adjusts generalization strategies according to the difficulty of blocking states, thereby improving scalability without compromising efficiency. A comprehensive evaluation demonstrates that EXCTG and DynAMic achieve significant scalability improvements, solving \EXCTGMORE\ and \DYNMORE\ more cases, respectively, compared to CTG generalization.
\end{abstract}

\section{Introduction}
IC3 \cite{IC3}, also known as PDR \cite{PDR}, is a prominent SAT-based model checking algorithm widely used in hardware formal verification. It efficiently searches for inductive invariants without unrolling the model. IC3 is distinguished by its completeness in comparison to BMC \cite{BMC} and its scalability compared to Interpolation-based Model Checking \cite{IMC} and K-Induction \cite{KINDUCTION}. IC3, widely recognized as a state-of-the-art algorithm, serves as the core engine for many efficient model checkers \cite{ABC,NUXMV}.

To verify a property, IC3 aims to identify inductive invariants derived from a sequence of frames $F_0 \dots F_k$ that over-approximate the set of reachable states. A key procedure in IC3 is generalization (also known as minimum-inductive clause, or MIC). Given an unsafe state represented as a cube, the goal of generalization is to expand it to include as many additional unreachable states as possible, thereby reducing the number of iterations. The standard algorithm \cite{IC3} adopts the down strategy \cite{Down}, which attempts to drop as many literals as possible.

The results of standard generalization can sometimes be suboptimal. For example, when trying to block a literal-dropped cube $cand$ in frame $F_i$, the process only checks whether $\lnot cand$ is inductive relative to $F_{i-1}$. If it is not, the attempt to directly block $cand$ is abandoned. However, if the predecessors of $cand$ can be blocked in $F_{i-1}$, then $cand$ may then be blockable in $F_i$. To overcome this limitation, CTG generalization \cite{CTG} has been proposed. This method aims to block counterexamples to generalization (CTG, which are also the predecessors of $cand$) when dropping literals fails. By attempting to block all predecessors of $cand$ in $F_{i-1}$, and if successful, $cand$ can then be blocked in $F_i$. This approach results in smaller cubes, blocks larger state spaces, and improves scalability compared to the standard method.

The results of CTG may sometimes still be suboptimal. Since it only considers blocking the predecessors of $cand$, if blocking the predecessors fails, it abandons directly blocking $cand$, even though the predecessors of $cand$'s predecessors could still be blocked. To address this issue, we propose EXCTG, an extension of CTG. Similar to CTG, when literal dropping fails, it attempts to block the CTG. However, if blocking the CTG also fails, EXCTG tries to block the predecessors of the CTG, leading to better generalization.

While CTG and EXCTG provide better generalization, they also introduce higher computational overhead, as they require significantly more SAT queries than the standard method. Current IC3 implementations typically adopt a single strategy and set of parameters applied across all generalization processes. Using the standard approach may lead to insufficient generalization, reducing scalability. Conversely, opting for CTG or EXCTG can increase generalization overhead, and in some simpler cases where such strong strategies are unnecessary, performance may actually suffer. Finding an appropriate balance between generalization quality and computational overhead is challenging with a static strategy. To mitigate this, we propose DynAMic (Dynamic Adjustment of MIC strategies), which measures the difficulty of blocking a state based on the number of blocking attempts and dynamically adjusts the generalization strategy and parameters according to this difficulty. For states that are easy to block, it uses the lightweight standard strategy to reduce overhead. For more challenging states, it applies more effective generalization strategies, such as CTG or EXCTG, depending on the difficulty.

We conducted a comprehensive evaluation, and the results show that our proposed EXCTG and DynAMic solved \EXCTGMORE\ and \DYNMORE\ more cases, respectively, compared to CTG generalization.

\section{Preliminaries}
\label{Sec:Preliminaries}
We use notations such as $x, y$ for Boolean variables, and $X, Y$ for sets of Boolean variables. The terms $x$ and $\lnot x$ are referred to as literals. Cube is conjunction of literals, while clause is disjunction of literals. A Boolean formula in Conjunctive Normal Form (CNF) is a conjunction of clauses. It is often convenient to treat a clause or a cube as a set of literals. For instance, given a clause $c$, and a literal $l$, we write $l \in c$ to indicate that $l$ occurs in $c$.

A transition system, denoted as $S$, can be defined as a tuple $\langle X, Y, I, T\rangle$. Here, $X$ and $X'$ represent the sets of state variables in the current state and the next state respectively, while $Y$ represents the set of input variables. The Boolean formula $I(X)$ represents the initial states, and $T(X, Y, X')$ describes the transition relation. State $s_1$ is a predecessor of state $s_2$ iff $(s_1, s_2')$ is an assignment of $T$ ($(s_1, s_2') \models T$). A safety property $P(X)$ is a Boolean formula over $X$. A system $S$ satisfies $P$ iff all reachable states of $S$ satisfy $P$.

\begin{algorithm}[!t]
\caption{Overview of IC3}
\label{alg:ic3}
\begin{algorithmic}[1]
\Function{$relind$}{cube $c$, frame $i$}

\Comment{Is clause $\lnot c$ inductive relative to $F_i$?}
    \State \Return $\lnot sat(F_i\land \lnot c \land T \land c')$
\EndFunction
\\
\Function{$get\_predecessor$()}{}
    \State $model \coloneqq get\_model()$
    \Comment{assignment of last SAT call}
    \State \Return $\{l \in model \mid var(l) \in X\}$
\EndFunction
\\
\Function{$block$}{cube $c$, frame $i$}
    \If{$i = 0$}
        \State \Return $false$
    \EndIf
    \While{$\lnot relind(c, i-1)$}
        \State $p \coloneqq get\_predecessor()$
        \If{$\lnot block(p, i-1)$}
            \State \Return $false$
        \EndIf
    \EndWhile
    \State // different strategy configurations
    \If{use\_CTG}
        \State $gen \coloneqq ctg\_generalize(c, i-1, $ CTG\_LV$)$
    \Else
        \State  $gen \coloneqq standard\_generalize(c, i-1)$
    \EndIf
    \State $F_j \coloneqq F_j \cup \{\lnot gen\}, 1\leq j \leq i$
    \State \Return $true$
\EndFunction
\\
\Function{$propagate$}{frame $k$} \label{ic3:propagate}
    \For{$1 \leq i < k$}
        \For{each $c \in F_i \setminus F_{i+1}$} 
            \If{$relind(\lnot c, i)$}
                \State $F_{i+1} \coloneqq F_{i+1} \cup \{c\}$
            \EndIf
        \EndFor
        \If{$F_i = F_{i+1}$}
            \State \Return $true$
        \EndIf
    \EndFor
    \State \Return $false$
\EndFunction
\\
\Procedure{IC3}{$I,T,P$}
    \State $F_0 \coloneqq I,k \coloneqq 1,F_k \coloneqq \top$
    \While{$true$}
        \While{$sat(F_k\land \lnot P)$}
            \State $c \coloneqq get\_model()$
            \If{$\lnot block(c, k)$}
                \State \Return $unsafe$
            \EndIf
        \EndWhile
        \State $k \coloneqq k + 1,F_k \coloneqq \top$
        \If{$propagate(k)$}
            \State \Return $safe$
        \EndIf
    \EndWhile
\EndProcedure
\end{algorithmic}
\end{algorithm}

\begin{algorithm}
\caption{Standard Generalization}
\label{alg:StandardGeneralization}
\begin{algorithmic}[1]
\Function{$down$}{cube \textbf{ref} $c$, frame $i$}
    \While{$true$}
        \If{$I \land c \neq \bot$} \label{algline:DownInit}
            \State \Return $false$
        \EndIf
        \If{$relind(c, i)$}
            \State \Return $true$
        \EndIf
        \State $p \coloneqq get\_predecessor()$
        \State $c \coloneqq c \cap p$ \Comment{common literals in $c$ and $p$} \label{algline:DownCommonLit}
    \EndWhile
\EndFunction
\\
\Function{$standard\_generalize$}{cube $c$, frame $i$}
    \For{each $l \in c$}
        \State $cand \coloneqq c \setminus \{l\}$
        \If{$down(cand, i)$}
            \State $c \coloneqq cand$
        \EndIf
    \EndFor
    \State \Return $c$
\EndFunction
\end{algorithmic}
\end{algorithm}

\begin{algorithm}
\caption{CTG Generalization}
\label{alg:CTGGeneralization}
\begin{algorithmic}[1]
\Function{$ctg\_down$}{cube \textbf{ref} $c$, frame $i$, ctg\_level $cl$}
    \State $num\_ctg \coloneqq 0$
    \While{$true$}
        \If{$I \land c \neq \bot$}
            \State \Return $false$
        \EndIf
        \If{$relind(c, i)$}
            \State \Return $true$
        \EndIf
        \State $p \coloneqq get\_predecessor()$
        \If{$cl > 0$ and $num\_ctg <$ CTG\_MAX and $i > 0$ } \label{algline:CTGExceed}
            \If{$I \land c = \bot$ and $relind(p, i-1)$} \label{algline:CTGBlockPred}
                \State $gen \coloneqq ctg\_generalize(p, i-1, cl-1)$ \label{algline:CTGGenPred}
                \State $F_j \coloneqq F_j \cup \{\lnot gen\}, 1\leq j \leq i$
                \State $num\_ctg \coloneqq num\_ctg + 1$
                \State \textbf{continue}
            \EndIf
        \EndIf
        \State $num\_ctg \coloneqq 0$
        \State $c \coloneqq c \cap p$
    \EndWhile
\EndFunction
\\
\Function{$ctg\_generalize$}{cube $c$, frame $i$, ctg\_level $cl$}
    \For{each $l \in c$}
        \State $cand \coloneqq c \setminus \{l\}$
        \If{$ctg\_down(cand, i, cl)$}
            \State $c \coloneqq cand$
        \EndIf
    \EndFor
    \State \Return $c$
\EndFunction
\end{algorithmic}
\end{algorithm}

IC3 is a SAT-based model checking algorithm, which only needs to unroll the system at most once. It tries to prove that $S$ satisfies $P$ by finding an inductive invariant $INV(X)$ such that:
\begin{itemize}
\item $I(X) \Rightarrow INV(X)$
\item $INV(X) \land T(X,Y,X') \Rightarrow INV(X')$
\item $INV(X) \Rightarrow P(X)$
\end{itemize}

To achieve this objective, it maintains a monotone CNF sequence $F_0\ldots F_k$. Each \emph{frame} $F_i$ is a Boolean formular over $X$, which represents an over-approximation of the states reachable within $i$ steps. Each clause $c$ in $F_i$ is called \emph{lemma}. IC3 maintains the following invariant: 

\begin{itemize}
\item $F_0 = I$
\item $F_{i+1} \subseteq F_i$
\item $F_i \Rightarrow F_{i+1}$
\item $F_i \land T \Rightarrow F_{i+1}$
\item for $i < k, F_i \Rightarrow P$
\end{itemize}

A lemma $\lnot c$ ($c$ is a cube) is said to be \emph{inductive relative} to $F_i$ if, starting from the intersection of $F_i$ and $\lnot c$, all states reached in a single transition are located inside $\lnot c$. This can be expressed as a SAT query $sat(F_i \land \lnot c \land T \land c')$. If this query is satisfied, it indicates that $\lnot c$ is not inductive relative to $F_i$ because we can find a counterexample that starts from $F_i \land \lnot c$ and transitions outside of $\lnot c$. If lemma $\lnot c$ is inductive relative to $F_i$, it can be also said that cube $c$ is blocked in $F_{i+1}$. If we want to block the cube $c$ in $F_{i+1}$, we need to prove that $\lnot c$ is inductive relative to $F_{i}$.

Algorithm \ref{alg:ic3}, \ref{alg:StandardGeneralization} and \ref{alg:CTGGeneralization} provide an overview of the IC3 algorithm. The \textbf{ref} keyword in the function parameter indicates that it is passed by reference (\textbf{\&} in C++). This algorithm incrementally constructs frames by iteratively performing the blocking phase and the propagation phase. During the blocking phase, it focuses on making $F_k \Rightarrow P$. It iteratively get a cube $c$ such that $c \models \lnot P$, and block it recursively. This process involves attempting to block the cube's predecessors if it cannot be blocked directly. It continues until the initial states cannot be blocked, indicating that $\lnot P$ can be reached from the initial states in $k$ transitions thus violating the property. In cases where a cube can be confirmed as blocked, IC3 proceeds to enlarge the set of blocked states through a process called generalization. This involves dropping variables and ensuring that the resulting clause remains relative inductive, with the objective of obtaining a minimal inductive clause. The propagation phase tries to push lemmas to the top frame. If a lemma $c$ in $F_i \setminus F_{i+1}$ is also inductive relative to $F_i$, then push it into $F_{i+1}$. During this process, if two consecutive frames become identical ($F_i = F_{i+1}$), then the inductive invariant is found and the safety of this model can be proofed.

\begin{figure}
    \centering
    \includegraphics[width=0.48\textwidth]{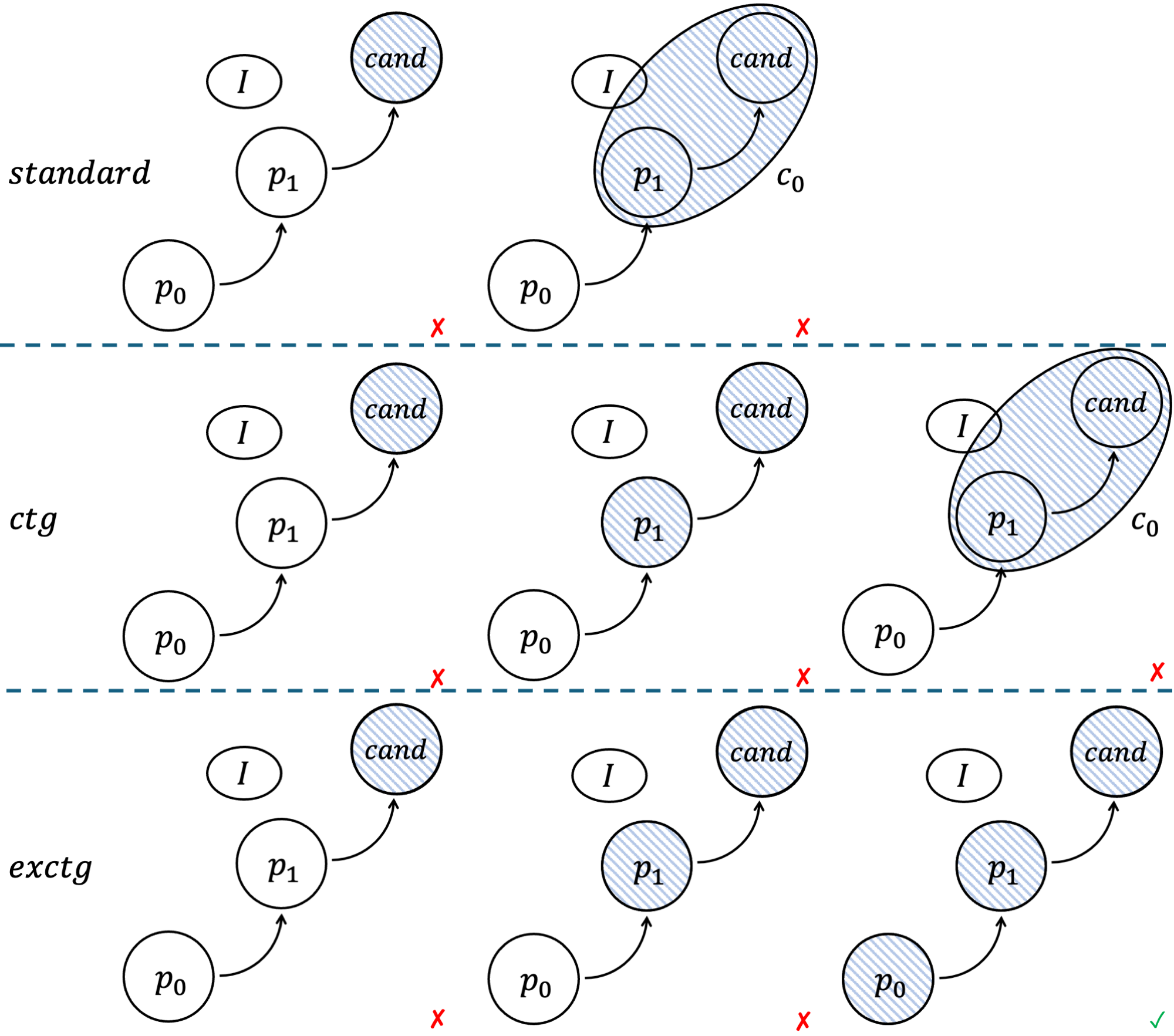}
    \caption{$p_0$, $p_1$, and $cand$ are cubes representing states, where $p_0$ is the predecessor of $p_1$, and $p_1$ is the predecessor of $cand$. $I$ represents the initial states. The cubes in shaded areas represent a set of states attempting to block. These diagrams illustrate the process of the different generalization strategies.}
    \label{fig:Strategy}
\end{figure}

To the best of our knowledge, there are currently two generalization strategies:
\begin{itemize}
    \item The standard generalization \cite{IC3,Down} uses \textit{down} to drop a literal, as shown in Algorithm \ref{alg:StandardGeneralization}. When trying to drop a literal $l$, it first attempts to block the cube $cand$ with $l$ removed. If successful, $l$ is dropped. If not, it then tries to block the cube that contains both $cand$ and the counterexample (Line \ref{algline:DownCommonLit}). For example in Fig. \ref{fig:Strategy}, the algorithm initially attempts to block $cand$, but this fails because $cand$ has a predecessor $p_1$, which has not yet been blocked. To block $cand$, $p_1$ must also be blocked. As a result, the algorithm tries to block $c_0$ (Line \ref{algline:DownCommonLit}), but this also fails because $c_0$ contains some initial states (Line \ref{algline:DownInit}). Consequently, $cand$ cannot be blocked, and literal dropping fails. We will refer to it as ‘Standard’ in the following sections.

    \item The CTG generalization \cite{CTG} uses $ctg\_down$ to drop a literal, as shown in Algorithm \ref{alg:CTGGeneralization} and Fig. \ref{fig:Strategy}. The key difference compared to $down$ is that if blocking $cand$ fails, it attempts to block the counterexample to generalization (CTG) of $cand$ ($cand$'s predecessor $p_1$) (Line \ref{algline:CTGBlockPred}). If the predecessor can be blocked, it will generalize it by recursively calling $ctg\_generalize$ (Line \ref{algline:CTGGenPred}), with a maximum recursion level $cl$. When $cl = 0$, $ctg\_generalize$ behaves the same as $standard\_generalize$. Therefore, $ctg\_generalize$ can be recursively called up to a maximum level of CTG\_LV. If all predecessors can be blocked, $cand$ will also be blocked. However, if blocking the predecessor fails ($p_1$ has a predecessor $p_0$), or if the number of predecessors that need to be blocked exceeds CTG\_MAX (Line \ref{algline:CTGExceed}), it will then attempt to block the cube $c_0$, which contains both $cand$ and its predecessors.
\end{itemize}

\section{Extended CTG Generalization}
As shown in Fig. \ref{fig:Strategy}, when blocking $cand$ fails, CTG attempts to block its predecessor, $p_1$. However, if blocking $p_1$ also fails, CTG abandons directly blocking $cand$ and instead tries to block a cube that contains both $cand$ and its predecessor. We attempt to put more effort into generalization: if blocking $p_1$ fails, we also attempt to block its predecessor, $p_0$. In Fig. \ref{fig:Strategy}, this succeeds because $p_0$ has no predecessor. As a result, $p_1$ can be blocked once $p_0$ is blocked, and $cand$ can then be successfully blocked. But if blocking $p_0$ fails, we continue by attempting to block the predecessor of the predecessor of $p_1$, and so on, to achieve better generalization.

\begin{algorithm}
\caption{EXCTG Generalization}
\label{alg:EXCTGGeneralization}
\begin{algorithmic}[1]
    \Function{\textcolor{deepblue}{$exctg\_block$}}{\textcolor{deepblue}{cube $c$, frame $i$, int \textbf{ref} $limit$, ctg\_level $cl$}}
    \If{$I \land c \neq \bot$}
            \State \Return $false$
    \EndIf
    \State $limit \coloneqq limit - 1$
    \If{$limit = 0$}
        \State \Return $false$
    \EndIf
    \While{true}
        \If{$\lnot relind(c, i-1)$}
            \State $p \coloneqq get\_predecessor()$
            \If{$\lnot exctg\_block(p, i-1, limit)$}
                \State \Return $false$
            \EndIf
        \Else
            \State $gen \coloneqq exctg\_generalize(p, i-1, cl)$
            \State $F_j \coloneqq F_j \cup \{\lnot gen\}, 1\leq j \leq i$
            \State \Return $true$
        \EndIf
    \EndWhile
\EndFunction
\\
\Function{$exctg\_down$}{cube \textbf{ref} $c$, frame $i$, ctg\_level $cl$}
    \State $num\_ctg \coloneqq 0$
    \While{$true$}
        \If{$I \land c \neq \bot$}
            \State \Return $false$
        \EndIf
        \If{$relind(c, i)$}
            \State \Return $true$
        \EndIf
        \State $p \coloneqq get\_predecessor()$
        \If{$cl > 0$ and $num\_ctg <$ CTG\_MAX and $i > 0$ }
            \textcolor{deepblue}{
                \If{$exctg\_block(i, p,$ EXCTG\_LIMIT$, cl - 1)$}
                    \State $num\_ctg \coloneqq num\_ctg + 1$
                    \State \textbf{continue}
                \EndIf
            }
        \EndIf
        \State $num\_ctg \coloneqq 0$
        \State $c \coloneqq c \cap p$
    \EndWhile
\EndFunction
\\
\Function{\textcolor{deepblue}{$exctg\_generalize$}}{\textcolor{deepblue}{cube $c$, frame $i$, ctg\_level $cl$}}
    \For{each $l \in c$}
        \State $cand \coloneqq c \setminus \{l\}$
        \If{$exctg\_down(cand, i, cl)$}
            \State $c \coloneqq cand$
        \EndIf
    \EndFor
    \State \Return $c$
\EndFunction
\end{algorithmic}
\end{algorithm}

We extend the CTG generalization (EXCTG), as presented in Algorithm \ref{alg:EXCTGGeneralization}. The modifications are highlighted in blue. When blocking cube $c$ fails, its predecessor $p$ is identified. EXCTG puts more effort into blocking $p$ by invoking the $exctg\_block$ function. The $exctg\_block$ function first attempts to block $p$. If this fails, the function recursively calls itself to block $p$'s predecessors. If blocking a predecessor of $p$ fails, it continues to block the predecessor’s predecessor, and so on. This process repeats until either all predecessors of $p$ are successfully blocked—thus allowing $p$ to be blocked—or the number of blocking attempts exceeds EXCTG\_LIMIT, at which point the function returns false.

\begin{figure}
    \centering
    \includegraphics[width=0.25\textwidth]{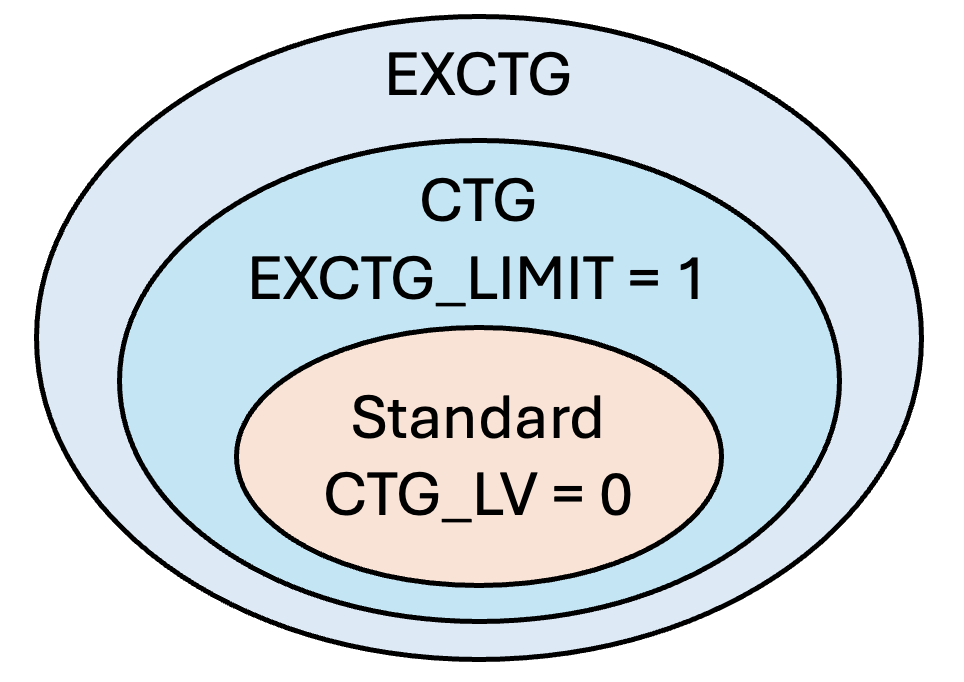}
    \caption{The relationships between Standard, CTG, and EXCTG.}
    \label{fig:Relation}
\end{figure}

The relationships between Standard, CTG, and EXCTG generalizations are illustrated in Fig. \ref{fig:Relation}. As shown, Standard is a special case of CTG (when CTG\_LV = 0), and CTG is a special case of EXCTG (when EXCTG\_LIMIT = 1).

\section{Dynamically Adjusting Generalization Strategies}

\begin{figure}
    \centering
    \includegraphics[width=0.48\textwidth]{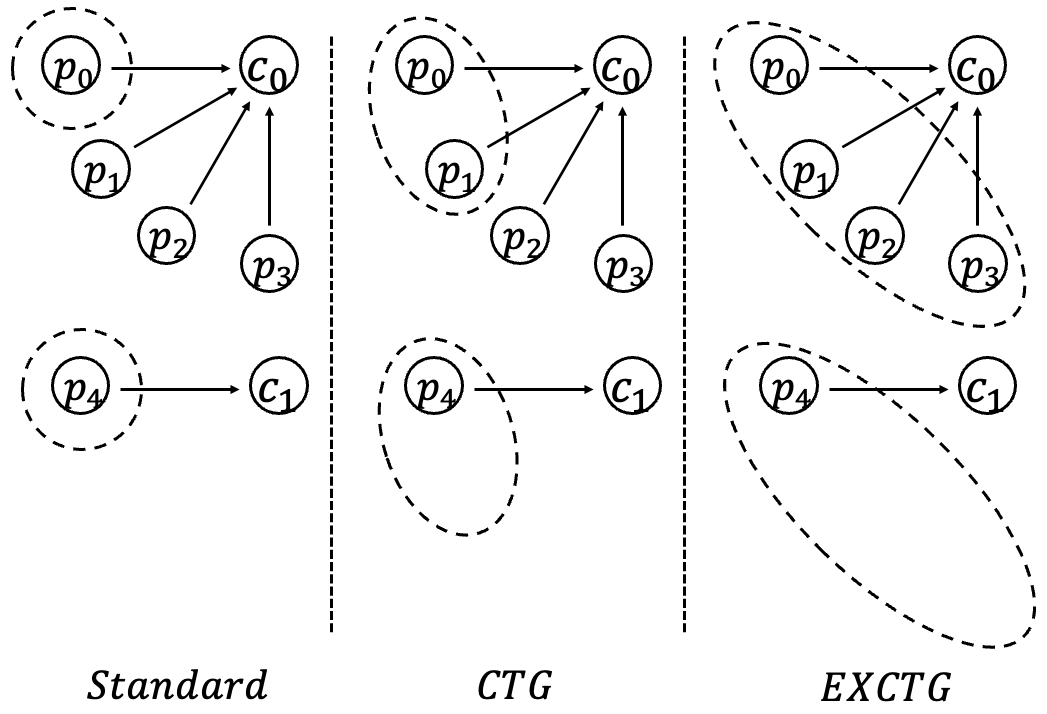}
    \caption{$c_0$ and $c_1$ represent bad states, and $p_i$ denote their predecessors. The dashed circle illustrates the state space generalized from $p_0$ or $p_4$ using different strategies.}
    \label{fig:DynAMic}
\end{figure}
While EXCTG provides better generalization results, its computational cost is significantly higher. Each time a literal is dropped, many more SAT calls are required compared to the standard approach. The generalization strategies: Standard, CTG, and EXCTG, produce progressively better results but also come with increased computational overhead. In the current implementations of the IC3 algorithm, the generalization strategy and its parameters are set at the beginning of the solving process and remain fixed throughout all subsequent generalization steps. However, the optimal strategies may vary depending on the specific bad states. For example, as shown in Fig. \ref{fig:DynAMic}, $p_0$ is better suited for generalization using EXCTG, as blocking $c_0$ requires all of its predecessors to be blocked. If the current generalization does not block all the predecessors, further blocking and generalization will need to continue in the next iteration. Conversely, $p_4$ is more efficiently handled by the Standard method, as CTG and EXCTG introduce more computational overhead.

It may be more effective to find a trade-off between generalization quality and computational overhead. Perhaps, by dynamically and adaptively selecting the appropriate generalization strategy for different states, we could better harness the strengths of each strategy. However, the key challenge lies in determining when each generalization strategy should be applied. Intuitively, the harder a state is to block, the more effort we should invest in generalizing its predecessors. We quantify the difficulty of blocking a state by the number of failed attempts. When blocking a state $c$ fails, we initially use the Standard strategy to generalize its predecessor. As the number of failed attempts to block $c$ increases, we gradually switch to CTG or EXCTG. In this way, if a state is easy to block, we use Standard to reduce generalization overhead. If a state is difficult to block, we gradually apply strategies with better generalization to avoid under-generalization.

\begin{algorithm}
\caption{DynAMic Generalization}
\label{alg:DynamicGeneralization}
\begin{algorithmic}[1]
\Function{\textcolor{deepblue}{$dyn\_generalize$}}{\textcolor{deepblue}{cube $c$, frame $i$, activity $act$}}
    \If{$act <$ CTG\_TH} 
        \State // standard generalization
        \State CTG\_LV $\coloneqq 0$
    \ElsIf{$act <$ EXCTG\_TH}
        \State // CTG generalization
        \State CTG\_LV $\coloneqq 1$
        \State EXCTG\_LIMIT $\coloneqq 1$
        \State CTG\_MAX $\coloneqq (act - $CTG\_TH$) / 10 + 2$
    \Else
        \State // EXCTG generalization
        \State CTG\_LV $\coloneqq 1$
        \State EXCTG\_MAX $\coloneqq 5$
        \State EXCTG\_LIMIT $\coloneqq (act - $EXCTG\_TH$)^{0.3} \cdot 2 + 5$\;
    \EndIf
    \State $c \coloneqq exctg\_generalize(c, i,$ CTG\_LV$)$
    \State \Return $c$
\EndFunction
\\
\Function{$block$}{cube $c$, frame $i$, \textcolor{deepblue}{successor\_activity $sact$}}
    \If{$i = 0$}
        \State \Return $false$
    \EndIf
    \State \textcolor{deepblue}{$act \coloneqq 0$}
    \While{$\lnot relind(c, i-1)$}
        \State \textcolor{deepblue}{$act \coloneqq act + 1$} \label{algline:DynMicActInc}
        \State $p \coloneqq get\_predecessor()$
        \If{\textcolor{deepblue}{$\lnot block(p, i-1, act)$}}
            \State \Return $false$
        \EndIf
    \EndWhile
    \State \textcolor{deepblue}{$gen \coloneqq dyn\_generalize(c, i-1, sact)$} \label{algline:DynGen}
    \State $F_j \coloneqq F_j \cup \{\lnot gen\}, 1\leq j \leq i$
    \State \Return $true$
\EndFunction
\end{algorithmic}
\end{algorithm}

We introduce a heuristic method called DynAMic (Dynamic Adjustment of MIC strategies), as shown in Algorithm \ref{alg:DynamicGeneralization}. When attempting to block a bad state $c$, an activity value $act$ is recorded, which increases after each failed blocking attempt (Line \ref{algline:DynMicActInc}), reflecting the difficulty of blocking $c$. If blocking $c$ fails, its predecessor $p$ is identified, and we attempt to block $p$. Once $p$ is successfully blocked, we generalize it using the function $dyn\_generalize$, which takes into account the $act$ of $p$’s successor, $c$ (Line \ref{algline:DynGen}).

The $dyn\_generalize$ function dynamically adjusts the generalization strategy and parameters based on $sact$. We predefined two thresholds: CTG\_TH and EXCTG\_TH. 
\begin{itemize}
    \item When $sact <$ CTG\_TH, we use the Standard strategy.
    \item When CTG\_TH $\leq sact <$ EXCTG\_TH, the CTG is used, and CTG\_MAX is adjusted linearly based on $sact$. As the difficulty of blocking $c$ increases, the maximum number of attempts to block CTG is raised accordingly.
    \item When $sact \geq$ EXCTG\_TH, the EXCTG strategy is applied, and EXCTG\_LIMIT is adjusted based on $sact$. As the difficulty of blocking $c$ increases, the maximum limits in EXCTG are adjusted upwards accordingly. However, since $sact$ can sometimes reach very large values, an excessively high EXCTG\_LIMIT could negatively impact performance. To mitigate this, the growth rate of EXCTG\_LIMIT is designed to gradually slow as $sact$ increases under the power function.
\end{itemize}

\section{Evaluation}
\label{Sec:Evaluation}

\subsection{Experiment Setup}
We implemented Standard, CTG, EXCTG, and DynAMic within the rIC3 model checker \cite{rIC3}, which is the 1st in the BV track of Hardware Model Checking Competition 2024 (HWMCC'24) \cite{HWMCC}. For CTG generalization, we set the parameters to CTG\_MAX $= 3$ and CTG\_LV $= 1$, following the original experiment in \cite{CTG}. For EXCTG, we used the same CTG parameters with the additional setting of EXCTG\_LIMIT $= 5$. For DynAMic, the parameters were set to CTG\_TH $= 10$ and EXCTG\_TH $= 40$. We also consider the IC3 implementations in the state-of-the-art system ABC \cite{ABC}, using the standard and CTG strategies with identical parameters.

We conducted all configurations using the complete benchmark suite from the HWMCC'19 and HWMCC'20, comprising a total of 536 cases in AIGER format, all under identical resource constraints: 16GB of memory and a 3600s time limit. The evaluations were performed on an AMD EPYC 7532 processor running at 2.4 GHz. To increase our confidence in the correctness of the results, all results from rIC3 are certified by certifaiger \cite{Certifaiger}. To ensure reproducibility, we have provided our experimental artifact \cite{Artifact}.

\subsection{Results}
Table \ref{tab:OveralResult} presents a summary of the overall results, showing the number of solved cases for each configuration, as well as the additional cases solved using rIC3-Standard as the baseline. It also displays the PAR-2 score, commonly used in SAT competitions. Fig. \ref{fig:Plot} shows the number of cases solved over time, while Fig. \ref{fig:Scatter} presents scatter plots comparing the solving times of different configurations. From these results, we make the following observations.

\subsubsection{Baseline} The comparison demonstrates that the rIC3 systems perform well compared to the state-of-the-art system, ABC \cite{ABC}. Therefore, it is appropriate to use rIC3-Standard as a baseline.

\subsubsection{EXCTG}
\begin{figure}[!t]
    \centering
    \includegraphics[width=0.48\textwidth]{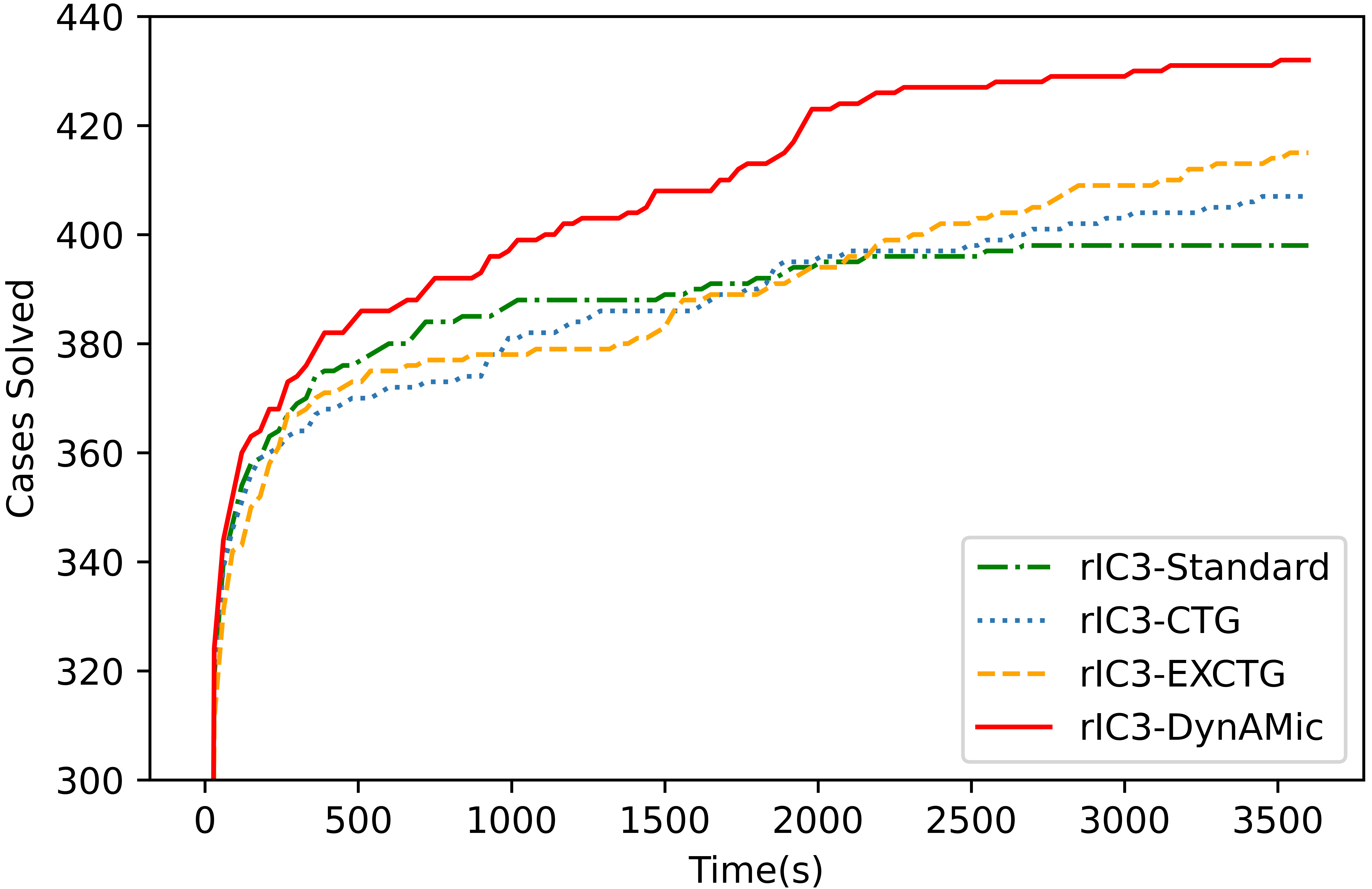}
    \caption{The number of cases solved by different configurations over time.}
    \label{fig:Plot}
\end{figure}

\begin{itemize}
    \item \textbf{Scalability.} CTG shows better scalability than Standard, consistent with the results in \cite{CTG}. Our proposed EXCTG solved \EXCTGMORE\ more cases than CTG, further highlighting its effectiveness in improving scalability.
    \item \textbf{Efficiency.} EXCTG exhibits lower efficiency compared to both Standard and CTG, as shown in Figure \ref{fig:Scatter} (b) and (c), with an increased solving time for most cases. This is because more SAT solver calls are made during each literal drop, leading to higher overhead. 
    \item As shown in Figure \ref{fig:Plot}, CTG initially solves fewer cases than Standard, but as time progresses, it surpasses Standard, consistent with the results reported in the original CTG paper \cite{CTG}. Similarly, EXCTG follows the same pattern. Due to EXCTG’s lower efficiency, it starts off slower, but its better scalability enables it to solve more cases over time.
\end{itemize}

\subsubsection{DynAMic} 
\begin{figure}[!t]
    \centering
    \includegraphics[width=0.48\textwidth]{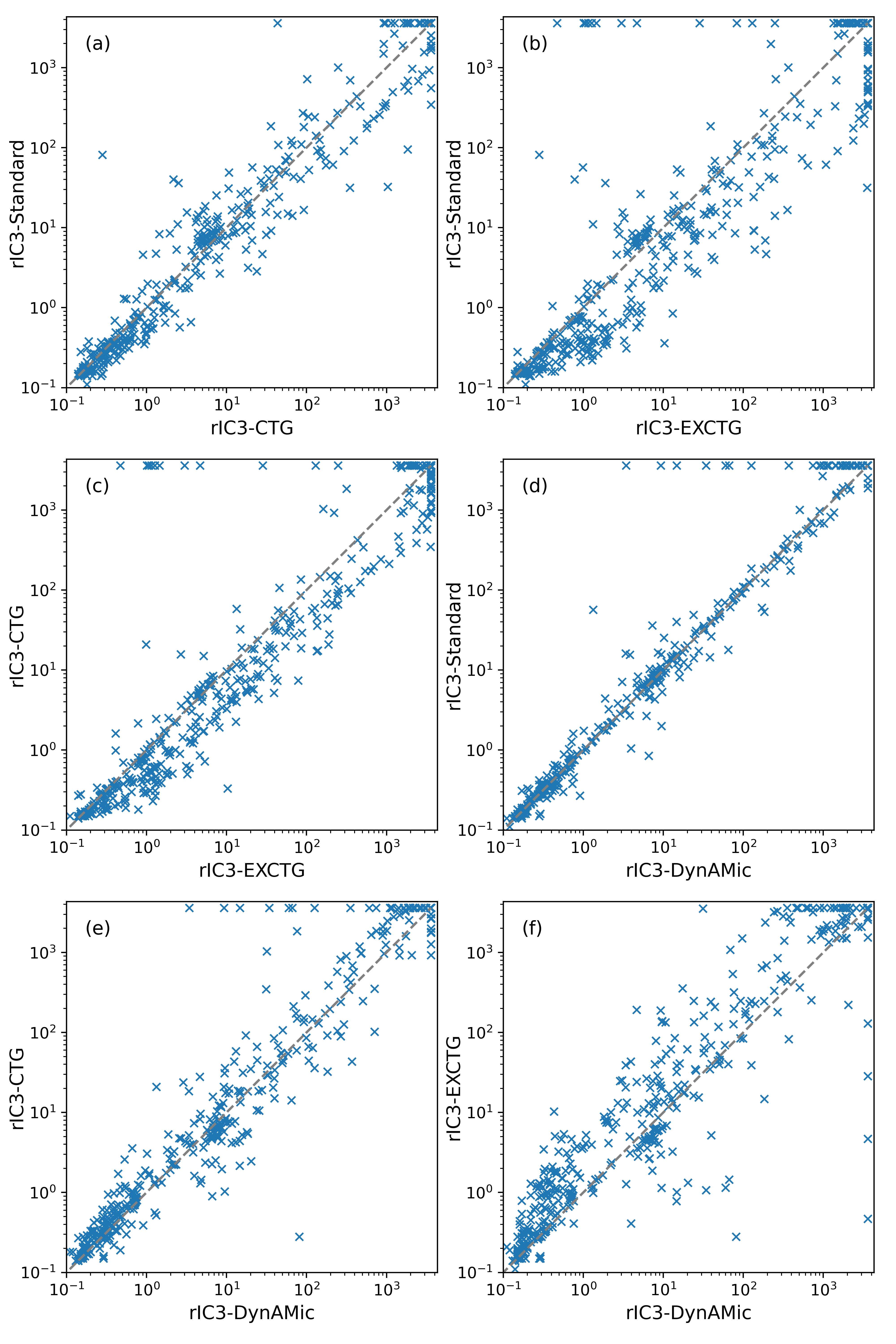}
    \caption{This plot compares the solving times (in seconds) between different configurations.}
    \label{fig:Scatter}
\end{figure}

\begin{itemize}
    \item \textbf{Scalability.} DynAMic demonstrates significant scalability improvements, solving \DYNMORE\ more cases than CTG and \DYNMORETOEXCTG\ more cases than EXCTG. This result highlights the effectiveness of dynamically adjusting strategies.
    \item \textbf{Efficiency.} Although DynAMic demonstrates significant improvements in scalability, its efficiency remains comparable to Standard and CTG, while exceeding EXCTG, as shown in Fig. \ref{fig:Scatter} (d), (e), and (f).
    
\end{itemize}

\begin{table}
\centering
\caption{Summary of Results}
\label{tab:OveralResult}
\begin{tabular}{c c c c c}
\hline
Configuration & \textbf{\#}Solved & $\Delta$ & PAR-2 \\
\hline
rIC3-Standard & 398 & 0 & 1922.54 \\
rIC3-CTG & 407 & +9 & 1866.83 \\
rIC3-EXCTG & 415 & +17 & 1802.63 \\
rIC3-DynAMic & \textcolor{red}{432} & \textcolor{red}{+34} & \textcolor{red}{1555.32} \\
\hline
ABC-PDR-Standard & 363 & 0 & 2405.96 \\
ABC-PDR-CTG & 369 & +6 & 2352.13 \\
\hline
\end{tabular}
\end{table}

\section{Related Work}
\label{Sec:RelatedWork}
Generalization is a critical component of the IC3 algorithm, and numerous efforts have focused on enhancing it.

The original IC3 algorithm employs down \cite{Down} to drop literals, significantly reducing the number of iterations. Building on this, CTG generalization \cite{CTG} aims to block counterexamples when literal dropping fails, achieving a more effective generalization. Details of both strategies are provided in Section \ref{Sec:Preliminaries}. We extend CTG by attempting to block the predecessors of counterexamples, which further enhances generalization. Additionally, we achieve a balance between generalization quality and computational overhead through dynamic strategies.

Some works have enhanced generalization while still utilizing either the Standard or CTG. In \cite{PredictingLemmas}, the authors aimed to predict the outcome before generalization, potentially reducing overhead if successful. The algorithm in \cite{CAV23} drops literals that do not appear in any subsumed lemmas from the previous frame, increasing the likelihood of propagating to the next frame. These two methods are not in conflict with our proposed methods and can be used simultaneously.

\section{Conclusion}
\label{Sec:Conclusion}
In this paper, we present a novel generalization strategy called EXCTG, which extends CTG. Building on both existing approaches and EXCTG, we introduce DynAMic, a heuristic method that dynamically adjusts MIC strategies and parameters. Our evaluation demonstrates that these proposed approaches lead to significant improvements in scalability.

\bibliographystyle{IEEEtran}
\bibliography{cites}
\end{document}